\documentclass[12pt,a4paper]{elsarticle} 

\usepackage[left=2cm,right=2cm,top=2cm,bottom=2cm]{geometry}
\usepackage[utf8]{inputenc}
\usepackage[english]{babel}
\usepackage{amsmath}
\usepackage{amsfonts}
\usepackage{amssymb}
\usepackage{graphicx}
\usepackage{physics}
\usepackage{indentfirst}

\usepackage[colorlinks=true, citecolor=blue, urlcolor=blue, linkcolor = blue]{hyperref}

\usepackage{float}
\usepackage[caption = false]{subfig}
\newcommand{\sign}{\operatorname{sign}}

\begin{document}

\begin{frontmatter}

\title{"Bound luminosity" state in the extended Dicke model}

\author[inst1]{S. S. Seidov}
\author[inst1]{S. I. Mukhin}

\affiliation[inst1]{organization={Theoretical Physics and Quantum Technologies Department, NUST "MISIS"}, city={Moscow},country={Russia}}

\begin{abstract}
The extended Dicke model describes interaction of the single--mode electromagnetic resonator with an ensemble of interacting two--level systems. In this paper we obtain quasiclassical equations of motion of the extended Dicke model. For certain initial conditions and range of parameters the equations of motion can be solved analytically via Jacobi elliptic functions. The solution is a "bound luminosity" state, which was described by the authors previously for ordinary Dicke model and now is generalized for the case of the extended Dicke model. In this state the periodic beatings of the electromagnetic field occur in the microwave cavity filled with the ensemble of two--level systems. At the beginning of the time period the energy is stored in the electromagnetic field in the cavity, then it is absorbed by the ensemble of two--level systems, being afterwards released back to the cavity in the end of the period. Also the chaotic properties of the semiclassical model are investigated numerically.
\end{abstract}
\end{frontmatter}

\section{Introduction}
In this paper we study the quasiclassical dynamics of the extended Dicke model, in the development of our previous work for ordinary Dicke model \cite{Seidov_JETP}. The Dicke model describes interaction of the single--mode electromagnetic resonator with an ensemble of two--level systems (TLSs) \cite{Dicke, Garraway, Hepp}. It was demonstrated by a straightforward derivation \cite{Rabl_2016} that extended Dicke model arises e.g. in the case of a single-mode $LC$ (with inductance $L$ and capacitance $C$) resonator capacitively coupled via an additional capacitance $C_g$ to array of  Cooper pair boxes with capacitances $C_q$ and Josephson energies $E_J$. An alternative derivation for the same system leading to the extended Dicke model was also made via the canonical phase shift of the gauge invariant Josephson phases \cite{Shnirman,Stroud_2001,Stroud_2002,Mukhin}. Both approaches led to the extended Dicke model Hamiltonian of the type presented below in Eq. (\ref{eq:HD}), but with $\varepsilon=0$. Namely, starting from Hamiltonian:

\begin{equation}
H = \frac{\hat p^2 + \omega^2 \hat q^2}{2} + E_C \sum_{i=1}^N \hat n_i^2 - E_J \sum_{i=1}^N \cos \left(\hat \phi_i - \displaystyle{\frac{g}{\hbar}} \hat{q} \right),\label{H}
\end{equation}
where $\hat p$ and $\hat q$ are operators of momentum and coordinate of the photonic oscillator respectively, with Josephson junction energy $E_J$, and its capacitance $C_q$ entering the capacitive energy of the junction $E_C \hat n^2 =(2e)^2/(2C_q)\hat n^2$, $\hat n$ is an operator which represents the difference between the number of Cooper pairs on the two superconducting islands which form the junction, so that $2 e \hat n_i$ is the charge of the junction, $\hat \phi$ is the Josephson phase operator shifted in Eq. (\ref{H}) by a gauge term $ g\hat q$, where the coupling constant is \cite{Stroud_2002}:
 \begin{equation}
g={2 el\sqrt{4\pi}}/(\sqrt{V}),\label{g}
\end{equation}
\noindent
and $l$ is the effective thickness across a Josephson junction, $V$ being the cavity volume.  Hence, we deal with the two mutually commuting sets of the conjugate variables: $\left[\hat{p},\hat{q}\right]=-\mathrm{i\hbar}$ and $\left[\hat{n}_i,\hat{\phi}_i\right]=- \mathrm{i}$. Next, following \cite{Stroud_2002} one makes a canonical transformation: 
\begin{align}
\hat{\phi}'_i=\hat{\phi}_i -\displaystyle{\frac{g}{\hbar}} \hat{q}\,  \;\;\; \text{and} \;\;\; \hat{n}'_i=\hat{n}_i \, \label{canonical_transform_JJ}
\end{align}
for the Josephson variables and:
\begin{align}
\hat{p}'=\hat{p} + g \sum_{i=1}^N \hat{n}_i\,  \;\;\; \text{and} \;\;\; \hat{q}'=\hat{q} \, \label{canonical_transform_pq}
\end{align}
\noindent for photonic variables. This transformation keeps intact commutation relation $\left[\hat{p}',\hat{\phi}'_i\right]=0$ as well as commutation relations between all the other operators. The Hamiltonian Eq. (\ref{H}) becomes:
\begin{align}
\hat{H} =  \frac{1}{2}\left(\hat{p}^2+\omega^2 \hat{q}^2 \right) - g \hat{p} \sum_{i=1}^N \hat{n}_i +\frac{g^2}{2} \left(\sum_{i=1}^N \hat{n}_i\right)^2 + \sum_{i=1}^N\left( E_C \, \hat{n}_i^2 - E_J \cos \hat{\phi}_i \right) \, \label{H_transformed},
\end{align}
\noindent where the primes in the new variables are omitted. Thus, the infinitely coordinated interaction term $\propto g^2$ has appeared in the canonically transformed hamiltonian (\ref{H_transformed}). Finally, in the limit $E_C\gg E_J$ charge and phase difference operators $\hat{n}_i$ and $\cos \hat{\phi}_i$ are projected in the two lowest energy levels approximation on the pseudo spin $1/2$ operators $\hat{s}_i^z$ and $\hat{s}_i^x$ correspondingly, where $2\hat{s}^\alpha = \hat{\sigma}^\alpha$ are the Pauli matrices. Then, after a unitary transformation interchanging the pseudo spin operators as $\hat{s}_i^z\rightarrow -\hat{s}_i^y, \hat{s}_i^x\rightarrow \hat{s}_i^z, \hat{s}_i^y\rightarrow -\hat{s}_i^x$, one finds the same expression as in \cite{Rabl_2016}: 

\begin{align}
\hat{H} =  \frac{1}{2}\left(\hat{p}^2+\omega^2 \hat{q}^2 \right) + {g} \hat{p} \sum_{i=1}^N \hat{s}^y_i  - E_J\sum_{i=1}^N \hat{s}^z_i +\frac{{g}^2}{2} \left( \sum_{i=1}^N \hat{s}^y_i\right)^2  \, \label{H_CPB}.
\end{align}

\noindent Thus, initial Hamiltonian (\ref{H_transformed}) reduces to the extended Dicke model spin-boson Hamiltonian presented below in Eq. (\ref{eq:HD}) with $\varepsilon=0$, \cite{Rabl_2016,Stroud_2002,Mukhin}. Finally, a direct dipole-dipole coupling between different Josephson junctions $i,j$ inside the resonant cavity can be taken into account by $\hat{H}_{dd}$ Hamiltonian:

\begin{eqnarray}
&D_{ij} =\dfrac{r_0^3}{4\pi}\dfrac{|\vec{r}_{ij}|^2-3(\vec{r}_{ij}\cdot \vec{e}_z)^2}{|\vec{r}_{ij}|^5},\quad \vec{r}_{ij}=\vec{r}_{i}-\vec{r}_{j}
\label{Dij},\\
&\hat{H}_{dd}=\dfrac{g^2}{2}\displaystyle\sum_{i, j}D_{ij}\hat{s}^y_i \hat{s}^y_j \rightarrow  \dfrac{\varepsilon g^2}{2} \hat {S}_y^2 ,\quad \hat{S}_y=\displaystyle\sum_{i}\hat{s}^y_i\label{HDfin}
\end{eqnarray}
\noindent where $D_{ij}$ is a dimensionless coupling parameter for the neighbouring dipoles separated by a distance $r_0$, and $\varepsilon\lessgtr 0$ describes ferroelectric/anti-ferroelectric dipole couplings \cite{Rabl_2018}. Hence, after adding $\hat{H}_{dd}$ from Eq. (\ref{HDfin}) to the Hamiltonian $\hat{H}$ in Eq. (\ref{H_CPB}) one arrives at the final expression in Eq. (\ref{eq:HD}) below. At a certain critical value of the coupling strength $g$ between the resonator and the  pseudo spin ensemble a superradiant or subradiant phase transition occurs \cite{Rabl_2016, Mukhin, Rabl_2018, Rabl,  Seidov_JOPA,  Brandes} depending on the $\varepsilon\leqslant 0$ or $\varepsilon > 0$ type of direct interaction between dipoles. The transition leads to appearance of the macroscopic photonic condensate in the resonator and consequent change of the quantum mechanical state of the ensemble of the two--level systems.  Previously we have described the "bound luminosity" state \cite{Seidov_JETP} appearing in quasiclassical dynamics of the ordinary Dicke model, i.e. the one without quadratic interaction term between two--level systems in the Hamiltonian Eq. (\ref{H_CPB}). The quadratic term disappears e.g. when there exists a ferroelectric coupling between the dipoles with $\varepsilon= -1$ in the extended Dicke model Hamiltonian , see Eq. (\ref{eq:HD}) below. We call "bound luminosity" state the periodic process of energy transfer between two--level systems and the superradiant photonic condensate in the cavity. Suppose that at some initial moment each of the two--level systems occupies its lowest energy level and all energy of the system is stored in the photonic condensate in the cavity. Then, the energy of the photonic condensate is transferred  coherently to the two--level systems, leading to decay of electromagnetic field in the cavity and transition of the two--level systems in their excited state. After that the energy will be again transferred back to the photonic condensate and the process will repeat. In this paper we obtain analytical expressions, describing these beatings in the case of the extended Dicke model.

The paper is structured as follows. First we consider the extended Dicke model Hamiltonian and obtain quasiclassical equations of motion of corresponding quantum observables. Obtained equations of motions are solved in a certain approximation and solutions describing "bound luminosity" state are found analytically. These solutions generalise our previous result, obtained for particular case of ordinary Dicke model \cite{Seidov_JETP}.   

\section{Dynamics of the extended Dicke model}
The extended Dicke model describes $N$ two--level systems (TLS) interacting with a single mode electromagnetic resonator. Each TLS is described by the spin--$1/2$ Pauli matrices $\hat \sigma^{x,y,z}_i$, and the electromagnetic wave in the resonator is described as an oscillator with coordinate $\hat q$ and momentum $\hat p$ assuming wave-length of the field is much greater then the inter-TLS distances. The Hamiltonian is: 
\begin{equation}\label{eq:HD}
\hat H = \frac{\hat p^2 + \omega^2 \hat q^2}{2} + g \hat p \hat S_y - \omega_0 \hat S_z + (1 + \varepsilon) \frac{g^2}{2} \hat S_y^2,
\end{equation} 
where $\omega$ is the resonator frequency, $g$ is the coupling constant, $\omega_0$ is the energy distance between energy levels of TLS, the last term  $\sim (1+\varepsilon)$ describes inter--spin interaction \cite{Mukhin,Rabl_2018}. The total spin projections are introduced as:
\begin{equation}
\hat S_{x,y,z} = \sum_{i=1}^N \hat \sigma_i^{x,y,z}.
\end{equation}
\noindent Expectation values of operators $\hat p$ and $\hat S_y$ determine the electric field in the resonator and the energy of the dipole moment of the TLS ensemble in this electric field:
\begin{equation}\label{eq:Ed}
\begin{aligned}
\vb E &={\bf{\hat{n}}} \sqrt\frac{4\pi}{V} \langle \hat p \rangle\\
\vb d& = -2 e l {\bf{\hat{e}}_y}\langle \hat S_y \rangle,\\
-\vb E\vb d &=g  \langle \hat p \rangle\langle \hat S_y \rangle.
\end{aligned}
\end{equation}
Here $V$ is the volume of the microwave cavity, $e$ is the charge of the electron, $l$ is the effective size of the dipole, ${\bf{\hat{n},\hat{e}}_y}$ are polarisations unit vectors.

\subsection{Quasiclassical equations of motion}
First we obtain the Heisenberg equations of motion for operators $\hat p$, $\hat q$, $\hat S_{x,y,z}$ via their commutator with the Hamiltonian:
\begin{equation}
\dot{\hat A} = i [\hat H, \hat A],\ \hat A = \hat p, \hat q, \hat S_{x,y,z}.
\end{equation} 
The following expressions arise due to $\hat S_y^2$ term in the Hamiltonian:
\begin{equation}\label{eq:SxSy}
\begin{aligned}
&[\hat S_z, \hat S_y^2] = i \hat S_x \hat S_y + i \hat S_y \hat S_x\\
&[\hat S_x, \hat S_y^2] = i \hat S_y \hat S_z + i \hat S_z \hat S_y.
\end{aligned}
\end{equation}
In quasiclassical approximation we replace the quantum--mechanical operators with real valued functions which commute with each other, thus the r.h.s in (\ref{eq:SxSy}) just turns into $2 S_x S_y$ and $2 S_z S_y$. This can be justified by noticing, that the commutator of two operators $[\hat S_i, \hat S_j] \sim \hbar$ and $\hbar \rightarrow 0$ in the quasiclassical limit. Finally, the quasiclassical equations of motion are
\begin{equation}\label{eq:Dynamics}
\begin{aligned}
&\dot S_z = -g p S_x - (1 + \varepsilon)g^2 S_x S_y\\
&\dot S_x = g p S_z + \omega_0 S_y + (1 + \varepsilon)g^2 S_y S_z\\
&\dot S_y = -\omega_0 S_x\\
&\dot p = -\omega^2 q\\
&\dot q =  p + g S_y.
\end{aligned}
\end{equation}

\noindent This approximation is valid for large total spin $S$ and accordingly large number $N$ of TLSs in the resonator. In addition, later we will consider the case of the superradiant phase in which a macroscopic photonic condensate emerges. This means that operators $\hat p$ and $\hat S_y$ acquire large averages, dynamics of which can be described quasiclassicaly.

\subsection{Stationary points}
The phase space of the system is a product $\mathbb{S}^2 \times \mathbb{R}^2$, where $\mathbb{S}^2$ is a sphere --- phase space of the spin, and $\mathbb{R}^2$ is the $p$--$q$ plane --- phase space of the oscillator. We introduce a vector in the phase--space 
\begin{equation}
\vb x = \begin{pmatrix}
S_x, & S_y, & S_z, & p, & q
\end{pmatrix}^T.
\end{equation}
The stationary points are the points in phase space where the derivatives of $p$, $q$, $S_{x,y,z}$ vanish. Thus to find them one needs to solve system (\ref{eq:Dynamics}) with left hand side equal to zero. When solving this system of equations, one should bear in mind the total spin conservation law:
\begin{equation}\label{eq:square}
S_x^2 + S_y^2 + S_z^2 = S^2+S\approx S^2, \;S \gg 1.
\end{equation} 
 \noindent The system has solutions
\begin{equation}
\begin{aligned}
S_x &= 0 & S_y &= 0 & p &= 0 & q &= 0\\
S_x &= 0 & S_z &= -\dfrac{\omega_0}{\varepsilon g^2} & p &= - g S_y & q &= 0.
\end{aligned}
\end{equation}
From total spin conservation law we find $S_z = \pm S$ for the first solution and $S_y = \pm \sqrt{S^2 - \omega_0^2/(\varepsilon^2 g^4)}$ for the second one. The resulting stationary points are
\begin{equation}\label{eq:fixed_points}
\begin{aligned}
&\vb x_\pm^\text{pole} = 
\begin{pmatrix}
0, & 0, & \pm S, & 0, & 0
\end{pmatrix}^T\\
&\vb x_\pm = \begin{pmatrix}
0, & \pm \sqrt{S^2 -\dfrac{\omega_0^2}{\varepsilon^2 g^4}}, & - \dfrac{\omega_0}{\varepsilon g^2}, & \mp g \sqrt{S^2 -\dfrac{\omega_0^2}{\varepsilon^2 g^4}}, & 0
\end{pmatrix}^T.
\end{aligned}
\end{equation}
The first two stationary points $\vb x_\pm^\text{pole}$ correspond to spin aligned along $z$--axis and the oscillator being at the origin of the $p$--$q$ plane. Other two stationary points $\vb x_\pm$ exist when
\begin{equation}\label{eq:gc}
S^2 - \dfrac{\omega_0^2}{\varepsilon^2 g^4} > 0 \Rightarrow g > g_c = \sqrt\frac{\omega_0}{|\varepsilon| S}.
\end{equation}
For $-1<\varepsilon <0$ these points correspond to superradiant phase in which the spin tends to align itself along the $y$ axis and non--zero average of photonic momentum $\hat p$ appears \cite{Seidov_JOPA}. The $y$ component of the spin at $\vb x_\pm$ can be written as:
\begin{equation}
S_y \eval_{\vb x_\pm} = \pm S \sqrt{1 - \frac{g_c^4}{g^4}}.
\end{equation}  
\noindent Together with the expression (\ref{eq:gc}) for the critical coupling constant and given that $p \eval_{\vb x_\pm} = - g S_y  \eval_{\vb x_\pm}$ this reproduces known results for the superradiant phase transition \cite{Mukhin,Rabl_2018,Seidov_JOPA,Brandes}. 

\subsubsection{Stability of fixed points}
The Hamiltonian, as a function of canonical variables, has extrema at the fixed points of the corresponding system of equations of motion. The stability of the fixed point is then determined by the type of the extremum i.e. by it being a maximum or a minimum. Thus, first we should express the spin variables in the Hamiltonian (\ref{eq:HD}) in terms of canonical variables. We choose those as $S_z$ and $\varphi$, where $\varphi$ is the angle of rotation around $z$--axis and $\{\varphi, S_z\} = 1$. Then the Hamiltonian is:
\begin{equation}\label{eq:H_Sz_phi}
H = \frac{p^2 + \omega^2 q^2}{2} + g p \sqrt{S^2 - S_z^2} \sin \varphi - \omega_0 S_z + (1 + \varepsilon) \frac{g^2}{2}\qty(S^2 - S_z^2) \sin^2 \varphi. 
\end{equation}
 The fixed points obtained from conditions $\partial_q H = 0$, $\partial_p H = 0$ and $\partial_\varphi H = 0$ are $q = 0$, $p = \pm g \sqrt{S^2 - S_z^2}$, $\varphi = \mp \pi/2$. These are the fixed points $\vb x_\pm$, but expressed via variables $S_z$ and $\varphi$. Next we substitute these values into the Hamiltonian (\ref{eq:H_Sz_phi}) in order to study its extrema structure with respect to variable $S_z$. The result is
\begin{equation}
H_\text{f.p.} = -\omega_0 S_z + \frac{\varepsilon g^2}{2}(S^2 - S_z^2). 
\end{equation}
It is convenient to introduce angle $\gamma$ of rotation in $z$--$y$ plane such that $S_z = S \cos \gamma$ and $S_y = S \sin \gamma$. Then finally we have to find the extrema of the function
\begin{equation}
U(\gamma) = -\omega_0 S \cos \gamma + \frac{\varepsilon g^2}{2} S^2 \sin^2 \gamma.
\end{equation} 

For $g < g_c$ the function has extrema at $\gamma_0 = 0 + 2 \pi n$ and $\gamma_\pi = \pi + 2 \pi n$, $n \in \mathbb{Z}$. These angles correspond to $S_z = \pm S$, i.e. the fixed points $\vb x_\pm^\text{pole}$. The points $\gamma_0$ are stable minimums and $\gamma_\pi$ are unstable maximums for all values of $\varepsilon$. The minimums $\gamma_0$ for which $S_z = S$ correspond to the normal phase of the Dicke model, stable for $g < g_c$. 

\begin{figure}
\subfloat[a) $\varepsilon < 0$]{\includegraphics[width = 0.49\textwidth]{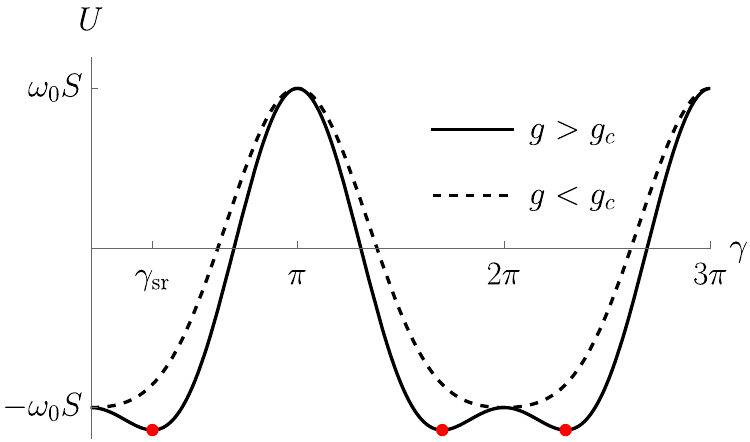}}
\subfloat[b) $\varepsilon > 0$]{\includegraphics[width = 0.49\textwidth]{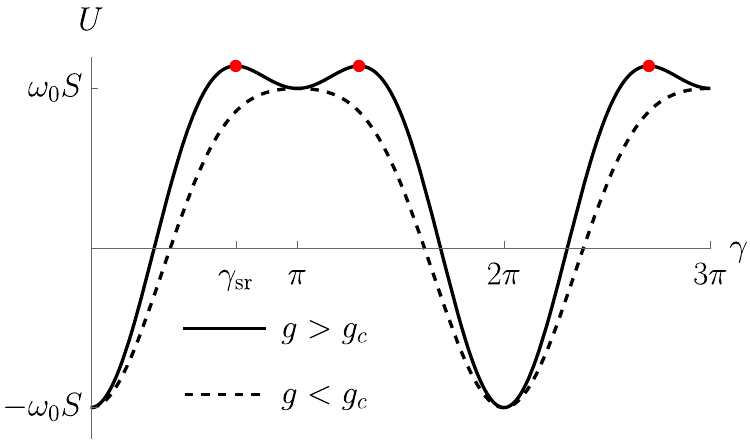}}
\caption{Plots of the function $U(\gamma)$, determining the stability of the Hamiltonian extrema and accordingly of the fixed points of the equations of motion (\ref{eq:Dynamics}). For $\varepsilon < 0$ (a) the extrema of U, corresponding to the superradiant phase (red dots) are minimums at $g > g_c$. For $\varepsilon > 0$ (b) they are maximums. Also the maximums at $\gamma = \pi n$ for $g > g_c$ and $\varepsilon < 0$ turn into minimums for $\varepsilon > 0$.
\label{fig:U}}
\end{figure}

For $g > g_c$ new extrema of $U$ appear for $\gamma_\text{sr}$ such that $\cos \gamma_\text{sr} = -\omega_0/(\varepsilon g^2 S)$. These points correspond to points $\vb x_\pm$ and describe the superradiant phase. They are minimums if $\varepsilon < 0$ and maximums if $\varepsilon > 0$. Thus we conclude, that the superradiant phase is stable for negative $\varepsilon$ and unstable for positive $\varepsilon$. Accordingly, the extrema $\gamma_0$ and $\gamma_\pi$ turn into (local)maximums if $\varepsilon < 0$ and into (local)minimums if $\varepsilon > 0$. The latter means, that the spin remains aligned along $z$--axis even for $g > g_c$, manifesting the stability of the subradiant phase for $\varepsilon > 0$. Plots of $U(\gamma)$ are presented in fig. \ref{fig:U}. 

\subsection{Bound luminosity solution for $p \approx -g S_y$}
Let us consider the case, when the displacement $q$ of the oscillator is small and $\dot q \approx 0, \dot p \approx 0$. Then from last two equations of (\ref{eq:Dynamics}) we obtain
\begin{equation}\label{eq:dqdp_zero}
\begin{aligned}
&q \approx 0\\
&p \approx - g S_y.
\end{aligned}
\end{equation}
Next we substitute these in the first three equation in (\ref{eq:Dynamics}) and obtain differential equations for spin projections:
\begin{equation}\label{eq:Dynamics_reduced}
\begin{aligned}
&\dot S_z = - \varepsilon g^2 S_x S_y\\
&\dot S_x =\omega_0 S_y + \varepsilon g^2 S_y S_z\\
&\dot S_y = -\omega_0 S_x.
\end{aligned}
\end{equation}
This system can be solved analytically. First we take the time--derivative of the last equation, express from it $\dot S_x$ and substitute in the second one:
\begin{equation}\label{eq:dd_Sy}
\ddot S_y = -\omega_0^2 S_y - \omega_0 \varepsilon g^2 S_y S_z.
\end{equation} 
Then in (\ref{eq:Dynamics_reduced}) we express $S_x$ from the last equation and substitute in the first one:
\begin{equation}
\dot S_z = \frac{\varepsilon g^2}{\omega_0} \dot S_y S_y =  \frac{\varepsilon g^2}{2\omega_0}\dv{t}\qty(S_y^2) \Rightarrow S_z = \frac{\varepsilon g^2}{2\omega_0} S_y^2 + C,\label{Sz}
\end{equation}
where $C$ is the conserved constant. Substituting (\ref{Sz}) in (\ref{eq:dd_Sy}) one obtains an equation for $S_y(t)$, which can be solved via Jacobi elliptic functions:
\begin{equation}\label{eq:dd_Sy_2}
\ddot S_y = -\omega_0^2 S_y - \frac{\varepsilon^2 g^4}{2} S_y \qty(S_y^2 + 2 C \frac{\omega_0}{\varepsilon g^2}).
\end{equation}

\subsubsection{Analysis using conservation laws}\label{sec:ES} 
\begin{figure}[h!!]
\subfloat[a)]{\includegraphics[width = 0.49\linewidth]{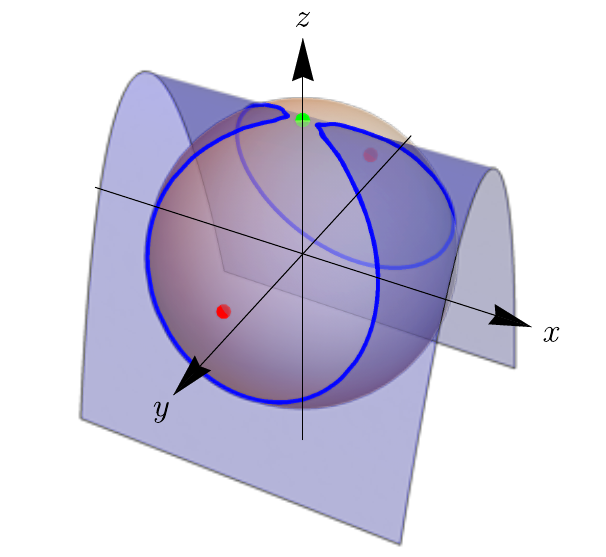}}
\subfloat[b)]{\includegraphics[width = 0.49\linewidth]{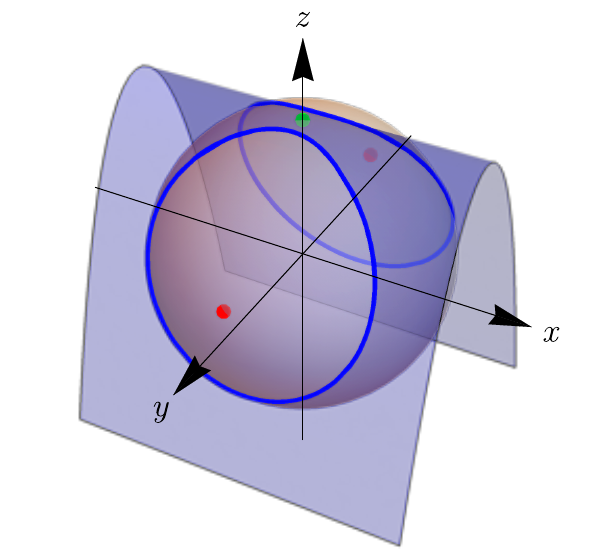}}
\caption{A single curve (a) and two separate curves (b) are the two types of phase portraits of the system (\ref{eq:Dynamics_reduced}) on the total spin sphere $S_x^2 + S_y^2 + S_z^2 = S^2$. The difference is due to the energy $E=-\omega_0C$ being greater in case (a) , then in case (b).  Blue surface is the surface of constant energy defined by Hamiltonian (\ref{eq:H_reduced}). Red dots --- stationary points $\vb x_\pm$, green dot --- stationary point $\vb x_+^\text{pole}$. The plots are made for $\varepsilon < 0$. For $\varepsilon > 0$ the plot of the constant energy surface is mirrored with respect to $x$--$y$ plane.}
\label{fig:sphere_intersec}
\end{figure}

Before solving the equation, it is instructive to use conservation laws for understanding properties of the resulting trajectories. If one substitutes $q = 0$ and $p = -g S_y$ in the Hamiltonian (\ref{eq:HD}), the following expression is obtained:
\begin{equation}\label{eq:H_reduced}
H \eval_{\substack{q = 0 \\ p = - g S_y}}= -\omega_0 S_z + \frac{\varepsilon}{2}g^2 S_y^2 = - \omega_0 C.
\end{equation}
So, there is a corresponding energy conservation law $E=-\omega_0 C$, and the second conservation law is the conservation of the total spin: $S_x^2 + S_y^2 + S_z^2 = S^2$. Thus the trajectory of the system should lay on the intersection of two surfaces defined by conservation laws, this is shown in fig. \ref{fig:sphere_intersec}. Similar results were obtained in \cite{Chinni_LMG}, when considering the Lipkin--Meshkov--Glick (LMG) model, which has the Hamiltonian with structure like equation (\ref{eq:H_reduced}). 

Depending on the energy, the intersection of two surfaces forms a connected curve or separates in to two distinct curves. In the superradiant regime at $g > g_c$ when the normal phase is unstable, connected trajectories correspond to meandering of the system between two stable superradiant phases, and separated trajectories --- to oscillations around a single stable point. 

\subsubsection{Solution of the equation of motion}
In this section we return to equation (\ref{eq:dd_Sy_2}). We rewrite it as
\begin{equation}\label{eq:dd_Sy_x}
\frac{\ddot S_y}{\omega_0 S} = -\frac{\omega_0}{2} \qty[\frac{g}{g_c}]^4 \qty(\frac{S_y}{S})^3 - \omega_0\qty(1 + \sign(\varepsilon) \qty[\frac{g}{g_c}]^2 \frac{C}{S}) \qty(\frac{S_y}{S}).
\end{equation}
This is the equation of motion of the particle with the mass $\omega_0^{-1}$ and coordinate $x = S_y/S$ in a quartic potential
\begin{equation}\label{eq:U}
U(x) = \frac{\omega_0}{8} \qty[\frac{g}{g_c}]^4 x^4 + \frac{\omega_0}{2} \qty(1 + \sign(\varepsilon) \qty[\frac{g}{g_c}]^2 \frac{C}{S})x^2.
\end{equation}
Thus, we can reduce the problem to a first order differential equation using total energy conservation law:
\begin{equation}
\tilde E = \frac{\dot x^2}{2 \omega_0} + U(x).
\end{equation}
It can be integrated using well known properties of double-periodic Jacobi functions \cite{Watson}:
\begin{equation}
\begin{aligned}
&x = k\frac{2\Omega}{\omega_0}\qty[\frac{g_c}{g}]^2 \operatorname{cn}\qty(\Omega t, k) \label{x}\\
&\dfrac{\Omega}{\omega_0} = \frac{g}{g_c}\left(\dfrac{\tilde E}{\omega_0^2 k^2(1-k^2)}\right)^{1/4}\ll 1,
\end{aligned}
\end{equation}
where energy $\tilde E$ is chosen small enough to guarantee inequality in (\ref{x}), which then justifies Eq. (\ref{eq:dqdp_zero}).  This inequality can be understood as condition for validity of adiabatic approximation in which the slow photonic condensate evolves on the background of fast TLSs in the ensemble. Also one should bear in mind that quasiclassical approximation for the $p, q$ variables is justified due to existence of the photonic condensate. Hence, solutions below are derived for $g > g_c$ and $\varepsilon < 0$. Now, for $S_x$, $S_y$ and $S_z$ we obtain: 

\begin{equation}\label{eq:sol_Sxyz}
\begin{aligned}
&S_y = x S \\
&S_z = S \qty[\frac{g_c}{g}]^2\left\{\qty[\frac{\Omega}{\omega_0}]^2 (2k^2{\operatorname{sn}\qty(\Omega t, k)}^2-1)+1\right\} \\
&S_x = -\frac{\dot S_y }{\omega_0} = 2Sk\qty[\frac{\Omega}{\omega_0}]^2\qty[\frac{g_c}{g}]^2{\operatorname{sn}\qty(\Omega t, k)}{\operatorname{dn}\qty(\Omega t, k)}.
\end{aligned}
\end{equation}
Here $\operatorname{cn}$, $\operatorname{sn}$ and $\operatorname{dn}$ are Jacobi elliptic functions. To find parameter $k$ one uses normalisation condition Eq. (\ref{eq:square}) and finds:

\begin{equation}\label{eq:k2}
k^2=\frac{1}{2}-O \left(\qty[\frac{\Omega}{\omega_0}]^2\right)
\end{equation}
\noindent It is worth to mention here, that according to Eq. (\ref{eq:gc}) the critical coupling strength $g_c\sim \sqrt{\omega_0}/\sqrt{|\varepsilon|S}$ is rather small for big TLS ensembles $S\sim N>>1$ at finite $|\varepsilon|$, and therefore a disorder in  $\omega_0$ for different member TLS would smear the point of quantum phase transition into superradiant state, but in the interval of small coupling strength values. Then, according to inequality condition  in Eq. (\ref{x}) this would lead to some narrowing of the interval of small enough energies $\tilde E$, that support  the ineqaulity, i.e. adiabatic approximation for the evolution of photonic degrees of freedom used in the above derivation that has led to results in Eq. (\ref{eq:sol_Sxyz}), but without qualitative change of the "bound luminosity" picture in fig. \ref{fig:Sz}.

\subsubsection{"Bound luminosity" state}
\begin{figure}[h!!]
\includegraphics[width=0.7\textwidth]{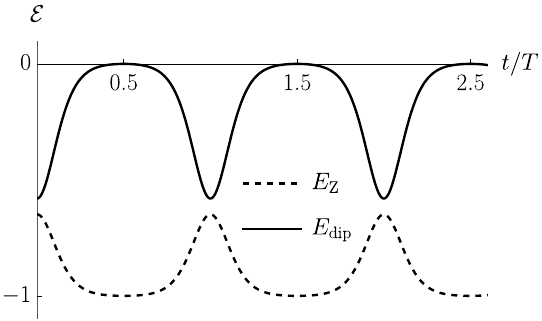}
\caption{Time dependencies of the Zeeman energy $E_\text{Z}$ of the spin subsystem (dashed line) and the dipole energy $E_\text{dip}$ (solid line) in the bound luminosity state with $g/g_c = 1.5$. Both are normalized such that the curves have the same scale on the plot: the energy $E_\text{Z}$ is plotted in units of $\omega_0 S$ and $E_\text{dip}$ --- in units of $gS^2$. Period $T=4K(k)/\Omega$, $K(k)$ is complete elliptic integral of the first kind, frequency $\Omega$ is given in Eq. (\ref{x}).}
\label{fig:Sz}
\end{figure}

As it follows from eq. (\ref{eq:Ed}), the electric field in the resonator is proportional to the momentum of the oscillator: $\vb{E} \sim p(t) $, and the dipole moment of the ensemble of the two--level systems is ${d} \sim - S_y$, also the momentum of the photonic oscillator $p(t) = - g S_y(t)$. Thus, given solutions (\ref{eq:sol_Sxyz}), one obtains the dipole energy  of the two--level ensemble in the electromagnetic field: $E_\text{dip} = -\vb{E} \vb{d} \sim p(t) S_y(t)$ and the Zeeman energy of the spin, associated with the ensemble of two--level systems $E_\text{Z} = -\omega_0 S_z(t)$.  Both energies $E_\text{dip}$ and $E_\text{Z}$ are plotted as functions of time in fig. \ref{fig:Sz}. 

From the plot one can clearly see what is the dynamical "bound luminosity" state. Suppose at the initial time the spin is aligned along positive direction of the $z$--axis, which minimizes its energy. Then it can absorb the energy of the photonic condensate, which leads to spin rotation towards negative direction of the $z$--axis. Accordingly, the photonic condensate decays. Next, the the TLS ensemble emits energy back to the resonator, reviving the condensate, that means the spin rotates into its initial direction, hence, the cycle repeats. Depending on initial conditions there are two types of trajectories as it is apparent from fig. \ref{fig:sphere_intersec}. Both of them are "bound luminosity" states as they share the same property of emergence and decay of photonic condensate in the cavity described above. The degree of energy transfer between the photonic and spin subsystems is defined by the distance at which the trajectory approaches the north pole of the $S^2$--sphere.

\subsection{Poincare sections and transition to chaos}
It was shown that there is a transition to quantum chaos in the Dicke model at critical coupling strength $g = g_c$ \cite{Brandes}. 

\begin{figure}[h!!]
\subfloat[a) $\varepsilon = -0.5$]{\includegraphics[width=0.7\textwidth]{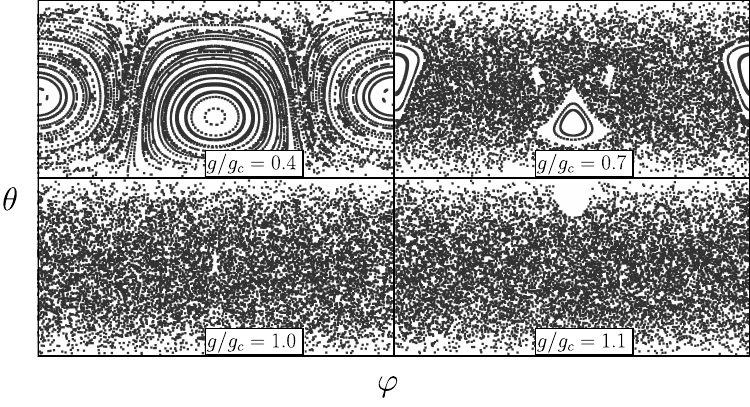}}\\
\subfloat[b) $\varepsilon = 1$]{\includegraphics[width=0.7\textwidth]{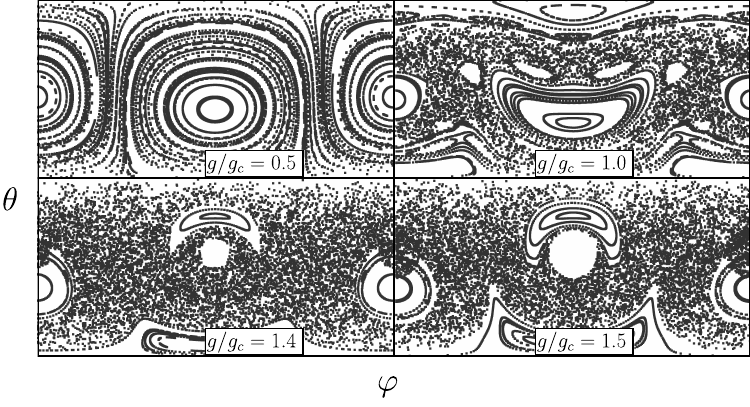}}
\caption{Poincare sections of the system (\ref{eq:Dynamics}): a) $\varepsilon = -0.5$ b) $\varepsilon = 1$. For both plots the rest of the parameters are chosen as: $\omega = 1$, $\omega_0 = 1$, $S = 5$ and $E = 10$. As coupling strength increases, regular trajectories are destroyed and more chaotic trajectories appear. At couplings higher than critical most of the phase space is chaotic.}
\label{fig:Poincare}
\end{figure}
To study this phenomenon in the extended Dicke model we numerically calculate the Poincare sections of the system (\ref{eq:Dynamics}). Although this is out of the scope of the present paper, Poincare sections in the phase space of classical systems are directly related to Wigner and Husimi phase space density distributions of its quantum counterpart \cite{Korsch_delocalization, Korsch_1981, Porter, Dalessio}. The Poincare section of a trajectory is constructed by choosing a plane in the phase space of the system and plotting the points at which the trajectory intersects the plane. For a regular (i.e. not chaotic) trajectory the points of the Poincare section lay on a curve. For a chaotic trajectory the section consists of randomly scattered points. In our case we define the section surface by $q = 0$ and $p = p(E)$, where $E$ is the total energy of the system. 
Thus, the section is obtained for the spin variables $S_{x,y,z}$. For convenience of making a plot we pass to the spherical angles $\theta$ and $\varphi$, defined as $\theta = \arccos(S_z/S)$, $\varphi = \arctan(S_y/S_x)$. Next we plot Poincare sections on a $\{\theta, \phi\}$ plane for trajectories with different initial conditions, see fig. \ref{fig:Poincare}. One can see, that at coupling constants well below critical the points are laying on the curves, thus manifesting regular dynamics. As coupling constant grows and approaches critical, more and more chaotic trajectories appear. For couplings above critical the chaotic trajectories fill the entire phase space. This was already shown for ordinary Dicke model ($\varepsilon = -1$) in ref. \cite{Brandes} and we have generalized this result for the extended Dicke model with arbitrary $\varepsilon$. The numerical results lead to a conclusion, that for $\varepsilon > 0$ the transition to chaos is suppressed, meaning that it requires higher ratio $g/g_c>1$ to happen.

\section{Conclusions}
In this work semiclassical equations of motion for extended Dicke model were obtained. In a certain approximation the class of solutions, describing "bound luminosity" state, is expressed analytically using Jacobi elliptic functions. In this state periodic beatings of the electromagnetic field in the cavity and of the energy of the ensemble of two--level systems occur. Namely, the energy is periodically transferred from the photonic condensate to the ensemble and back. This result generalizes described previously \cite{Seidov_JETP} "bound luminosity" state in the ordinary Dicke model. The corresponding Hamiltonian, generating this kind of dynamics, appeared to be an Lipkin--Meshkov--Glick model Hamiltonian. In addition, chaotic properties of the solutions of equations of motion were studied numerically by constructing Poincare sections of the system. The paper connects the areas of semiclassical dynamics in superradiant extended Dicke model with chaos in the superradiant state.
Namely, the paper presents analytic solutions of the semiclassical equations of motion of superradiant photonic condensate coupled to the array of two-level systems in extended Dicke model in the vicinity of the fixed points of the quasiclassical equations of motion. Numerical solutions of these equations demonstrate gradual transition to chaos in the Poincare sections of the phase space of the system depending on the strength of the coupling constant for different values of the direct coupling between two-level systems corresponding to the superradiant and subradiant phase transition phenomena in the extended Dicke model.

\section{Acknowledgments}
This work was supported by the federal academic leadership program "Priority 2030" (MISIS Strategic Project Quantum Internet)  grant No. K2-2022-025.  

\bibliographystyle{ieeetr}


\end{document}